%% file: paper_agnhi.tex
\documentclass[tighten,apj]{emulateapj}
\newcommand{\kms}{~km~s$^{-1}$}

\newcommand{\Hb}{{H$\beta$}}
\newcommand{\Ha}{{H$\alpha$}}	
\newcommand{\NII}{[N~{\footnotesize II}]} 
\newcommand{\SII}{[S~{\footnotesize II}]} 
\newcommand{\OIII}{[O~{\footnotesize III}]} 
\newcommand{\HI}{H\,{\footnotesize I}}
\newcommand{\HII}{H~{\footnotesize II}}

\newcommand{\MHI}{$\rm{M}_{\HI}$}
\newcommand{\MS}{$\rm{M}_{*}$}

\newcommand{\DMI}{$\Delta \rm{M_{\HI}}$}
\newcommand{\DBPT}{$d_{\rm BPT}$}
\newcommand{\Msun}{\rm{M}_{\odot}} 
\newcommand{\LOGMS}{\log{\rm{M_*/\Msun}}}
\newcommand{\LOGMHI}{\log{\rm{M_{\HI}/\Msun}}}

\input{m_star_m_hi_fit.tex} 
\input{catalog_stats.tex} 
\input{fiber.tex}

\begin{document}

\title{The Effect of AGN on the Global \HI\ Content of Isolated Low-Mass Galaxies}
\author{Jeremy D. Bradford \altaffilmark{1,4}, Marla C. Geha
  \altaffilmark{1}, Jenny E. Greene \altaffilmark{2}, Amy E. Reines
  \altaffilmark{3}, Claire M. Dickey \altaffilmark{1}}
\affil{$^1$ Astronomy Department, Yale University, New Haven, CT~06520, USA; \href{mailto:marla.geha@yale.edu}{marla.geha@yale.edu}}
\affil{$^2$ Department of Astrophysical Sciences, Princeton University, Princeton, NJ 08544, USA}
\affil{$^3$ Department of Physics, Montana State University, Bozeman, MT 59717, USA}
\altaffiltext{4}{NSF Graduate Research Fellow}

\begin{abstract}
\renewcommand{\thefootnote}{\fnsymbol{footnote}}

We investigate the global neutral hydrogen (\HI) content of isolated
galaxies selected from the SDSS spectroscopic survey with optical
evidence of Active Galactic Nuclei (AGN). Our sample includes galaxies
with unresolved \HI\ observations from the ALFALFA 70\% data release
combined with deeper \HI\ observations of low-mass galaxies with
$7.0 < \LOGMS < 9.5$. We examine the \HI\ masses of this sample using
the distance from the star-forming sequence on the \OIII/\Hb\ and
\NII/\Ha\ Baldwin Phillips Terlevich (BPT) diagram as a measurement of
AGN activity.  In agreement with previous studies, we find that, for
galaxies with $\LOGMS > 9.5$, AGN activity does not correlate with the
global \HI\ content. However, for galaxies with $9.2 < \LOGMS < 9.5$,
we identify a set of objects at large distances from the BPT
star-forming sequence and lower than expected \HI\ masses. This
gas-depleted sample is red in both g-r and NUV-r colors and compact
without distinguishable signs of star formation. This is surprising
because the vast majority of isolated galaxies in this stellar mass
regime are both star-forming and gas-rich.   These galaxies are greater than
1.5\,Mpc from any massive galaxy, ruling out environmental
processes as a source of the gas-depletion.  We suggest that either
black hole feedback or shocks from extremely bursty star formation
cause the emission lines and have destroyed or otherwise consumed the
cold gas.


\end{abstract}


\section{Introduction}
\label{sec_agnhi_intro}

Feedback from active galactic nuclei (AGN) is often invoked as a mechanism to regulate galaxy evolution \citep[e.g.,][]{Silk:1998up, Fabian:1999kw, Granato:2001eg, Granato:2004bm, DiMatteo:2005hl, Dashyan:2018}. AGN feedback, either alone or together with feedback from star formation, can have profound effects on galaxy evolution by heating, ionizing and expelling gas from the interstellar medium (ISM). However, important aspects of AGN feedback physics are poorly constrained and simulations are often tuned to reproduce observations and galaxy scaling relations \citep[e.g.,][]{Bower:2006fj, Somerville:2008ed, Gabor:2011ci, DeBuhr:2011iz, Schaye:2014gk, Genel:2014tj}.

Studies of the \HI\ content of active galaxies consistently produce results that AGN activity is not directly correlated with the global \HI\ gas mass of galaxies \citep[e.g.,][]{Ho:2008cca, Fabello:2011ig, Zhu:2015dg}. For example, \citet{Gereb:2015dy} find no relation between AGN \OIII\ luminosity and the global \HI\ content of massive galaxies, even when the AGN are split by red sequence and green valley. However, \citet{Lemonias:2014ik} find that high-mass galaxies with the largest gas fractions and the lowest \HI\ surface densities (and therefore the lowest star formation rate surface densities) are also likely to have signatures of AGN emission and large bulges. While AGN may affect cold gas locally, the above studies support the conclusion that AGN activity is not entirely responsible for the global \HI\ gas regulation in massive galaxies.

Recent studies have uncovered a large sample of low-mass galaxies with $\LOGMS < 9.5$ showing signatures of accreting black holes \citep[e.g.,][]{Reines:2013bp,Sartori:2105,Reines:2016il}. Low-mass galaxies offer an opportunity to examine AGN feedback processes in galaxies with shallower gravitational potential wells, where feedback processes may have a more profound effect on the global ISM than in high-mass galaxies. However, low-mass galaxies are strongly affected by environmental processes \citep[e.g.,][]{Geha:2012eu}. To clearly examine the properties of galaxies with evidence of AGN, we study isolated galaxies where the effects of environment are minimized.

\citet[][Paper\,I]{Bradford:2015km} showed that isolated galaxies with
$\LOGMS < 9.5$ cannot completely destroy their own cold gas
reservoirs.  Galaxies with $9.0 < \LOGMS < 9.5$ have gas fractions
greater than 20\% and galaxies with $\LOGMS < 9.0$ have gas fractions
greater than 30\%.   However, flux-limited surveys of \HI\ emission are, by definition, biased towards gas-rich
galaxies and may not contain low-mass galaxies that have
been significantly depleted of their \HI\ gas due to feedback
processes \citep{Koribalski:2004cv, Giovanelli:2005jt}.

In this paper, we ask whether the presence of an AGN significantly
affects the global \HI\ content of low mass galaxies ($\LOGMS < 9.5$).
We compliment the \HI\ observations from Paper I and data from the
ALFALFA \HI\ survey with new, deeper \HI\ observations of low-mass
galaxies with evidence of AGN activity.  In \S\ \ref{sec_agnhi_data},
we present the galaxy catalog, emission line diagnostic data,
\HI\ observations, and catalog statistics. We measure a new
\MHI\ versus \MS\ relation for all isolated ALFALFA galaxies and
explore emission line diagnostics by measuring the perpendicular
distance from the star forming region in \OIII/\Hb\ and
\NII/\Ha\ space. In \S\ \ref{sec_results}, we examine in detail
the \HI\ gas masses of isolated, low-mass galaxies with evidence of
gas-depletion due to AGN activity and compare these galaxies to a
gas-normal sample with similar positions on the \OIII/\Hb\ and
\NII/\Ha\ BPT diagram. We examine these populations and compare them
to the rest of the isolated sample. In \S\ \ref{sec_agnhi_discussion},
we discuss our results and posit what may cause the emission line
ratios and gas-depletion in this gas-depleted sample. In this work, we
adopt the following cosmological parameters: $\Omega_0 = 0.3$,
$\Omega_{\Lambda} = 0.7$, $H_0 = 70$~\kms\ (i.e., h = 0.7).

\section{Data and Methods} 
\label{sec_agnhi_data}

\subsection{Galaxy Catalog and Environment Definitions}
\label{subsec_agnhi_nsa}

We select galaxies from version 0.1.2 of the NASA Sloan Atlas (NSA)
catalog \footnote{\url{http://www.nsatlas.org}}
\citep{Blanton:2011dv}.   The NSA catalog is based on the
Sloan Digital Sky Survey DR8 (SDSS) \citep{Aihara:2011kj}. Stellar
masses are generated from the \citet{Blanton:2007kl} \texttt{kcorrect}
software and are calculated using a \citet{Chabrier:2003ki} IMF. We
assume a 0.2~dex uncertainty in \MS. We restrict our galaxy sample to
the NSA redshift between 0.002 and 0.055 and we apply several quality
cuts as detailed in Paper I.   These cuts ensure accurate photometry,
spectroscopy and heliocentric distance measurements.

Isolated galaxies have various definitions \citep{Haynes:1984el,
  Karachentsev:2011ju, Brough:2013df,
  Hirschmann:2013hv}. \citet{Geha:2012eu} find that low-mass galaxies
with $\LOGMS < 9.0$ located at least 1.5~Mpc from a massive galaxy are
always star-forming and isolated quenched galaxies with $9.0 < \LOGMS
< 9.5$ are extremely rare. We adopt this empirically motivated
isolation criterion for low-mass galaxies. Isolated low-mass galaxies
are therefore defined as $\LOGMS < 9.5$ with projected distances to a
massive host ($\log{M_*/M_{\odot} > 10}$) greater than
1.5~Mpc. Isolated high-mass galaxies are defined as in Paper~I with
$\LOGMS > 9.5$, a projected distance larger than 1.5~Mpc to the nearest galaxy more massive
than 0.5~dex, and a fifth nearest neighbor
surface density $\Sigma_{N} < 1~\rm{Mpc^{-2}}$. After restricting the
galaxies from the NSA catalog to isolated galaxies only, we obtain
\nnsaisoall\ isolated galaxies, of which \nnsaisohigh\ are high-mass
and \nnsaisolow\ are low-mass.

\subsection{\HI\ Observations}

\subsubsection{Our \HI\ Observations}
\label{subsubsec_agnhi_our_hi}

We have obtained new observations of \npostpaperone\ galaxies using
the Arecibo Observatory between March of 2015 and January of 2017.
The 21-cm observations used the same configuration detailed in
Paper I: L-band Wide receiver using 1024 resolution channels with a
bandwidth of 12.5 MHz and a velocity resolution of 2.6 km s$^{-1}$.
Position switched ON/ OFF observing and ON/OFF noise-diode calibration was used.
Our deeper integration times run from \mininttime\ to \maxinttime\
minutes on-source, with an average integration time of \meaninttime\
minutes.  We integrate until a detection has been made or we reach an
on-source exposure time of 30~minutes (although in a few cases the
exposure time is less due to observing constraints).

For this work, we re-reduce all \HI\ data from Paper I and our new
data in order to co-add all spectra from the same receiver
configuration and to maximize the number of usable observations.  Most
of the observations from Paper I focus on isolated galaxies with
$\LOGMS < 9$ and a Sersic axis ratio less than 0.65 in order to obtain
accurate kinematics from \HI\ line-widths.  Our new observations focus
mostly on the \citet{Reines:2013bp} ``dwarf" galaxy sample with signs
of AGN activity.  We have updated the Paper I \HI\ measurement
algorithm to salvage more noise-only spectra and spectra with unstable
baselines outside of the spectral window of interest.  The most
significant update to this algorithm is an adaptive Hanning smoothing
length (between 2.5 and 10 km/s) to optimize the measured emission
line parameters. Uncertainties in the properties of the \HI\ emission
line are calculated through the same bootstrap process as described in
Paper I.

We calculate upper limits for \nnondetectall\ non-detections in this
\HI\ sample. Non-detections are identified using a smoothed S/N cutoff
below 6 and verified by eye. For each non-detection, we estimate the
the \HI\ 20\% line-width using the stellar Tully-Fisher relation from \citet{Bradford:2016ia} and de-project this velocity based on the optical inclination. We then
calculate the expected \MHI\ using the stellar mass-to-gas mass
relation in Paper~I and  convert this gas mass to \HI\ flux. The
specific relationship between \MS\ and \MHI\ we use does not affect
our results, as the upper limits are always significantly less than
the predicted \MHI\ from this relation.  We create a synthetic
Gaussian emission line with this \HI\ line width and integrated
flux. We measure the rms of the observation using the un-smoothed
spectrum. We then iterate over this fiducial signal, add in a random
noise realization, and Hanning smooth this synthetic spectrum to a
resolution of 10 \kms. For each noise realization, we decrease the
integrated flux of the synthetic spectrum until we reach the smoothed
S/N cutoff of 6. When the S/N drops below 6, we perform 1000
additional noise realizations using this final synthetic emission
line. The median integrated flux of these random realizations where
the S/N is less than 6 is the final upper limit.

\input{tbl_new_hi_obs.tex}

We detail our new \HI\ observations, detections and non-detections, in Table \ref{tbl_new_hi_obs}. All of these galaxies have emission line measurements from \citet{Reines:2015dw}, which are described in \S\ \ref{subsec_agnhi_emldiag}. Note that some of these galaxies are isolated and some are not. For the duration of this work we will focus on isolated galaxies only, but we list these non-isolated galaxies for clarity.

\subsubsection{ALFALFA \HI\ Data}
\label{subsubsec_agnhi_alf_hi}

We compliment the \HI\ data described above with the publicly
available ALFALFA 70\% catalog, a blind, drift-scan survey
\citep{Haynes:2011en}. We match sources in the ALFALFA catalog using
their optical centers with ``detcodes" of 1 and 2 to NSA catalog
sources within 6'' and where the difference between systemic
velocities is less than 75\kms. We discard any matches to the ALFALFA
detcode 2 sources where the S/N is less than 6. For overlapping
observations between our data and the ALFALFA catalog, we
adopt the error-weighted average of the \HI\ measurements.

For galaxies without a detection that fall within the ALFALFA footprint, we calculate conservative upper limits by using the S/N calculation from \citet{Haynes:2011en} to back out the predicted flux for a S/N of 6:

\begin{equation}
S_{21} = \frac{6 \times W_{50} \times \sigma_{\rm rms}}{w_{\rm smo}^{1/2}},
\label{eq_agnhi_sn}
\end{equation}

\noindent with $S_{21}$ the \HI\ integrated flux density, $W_{50}$ the
50\% \HI\ line-width, $w_{\rm smo}~=~W_{50}/\Delta V_{\rm
  smooth}$[km/s], $V_{\rm smooth}$ is the resolution of the smoothed
spectrum and $\sigma_{\rm rms}$ the measured rms of the \HI\
spectrum. For this case, $\sigma_{\rm rms}
=$~\nalfalfarms~$~\rm{mJy}$, which is the mean rms of the ALFALFA
survey. We then calculate the \HI\ line widths using the stellar
Tully-Fisher relation from \citet{Bradford:2016ia} and upper limits as described in \S\,\ref{subsubsec_agnhi_our_hi}. 

While we are not guaranteed that this approach accurately represents
the upper limits of these galaxies, we find that these upper
limits match the ALFALFA Spaenhauer diagram (\MHI\ versus
heliocentric distance) and the \HI\ profile width versus integrated
flux density relations from ALFALFA (Figures 3 and 12 in
\citet{Haynes:2011en}. These results have motivated us to proceed with
these upper limits serving purely as a reference for our primary
analysis.

\subsection{The Isolated \MHI\ to \MS\ Relation}
\label{subsec_agnhi_hi_stellar_mass_relation}

\begin{figure}[t!]
\epsscale{1.23}
\plotone{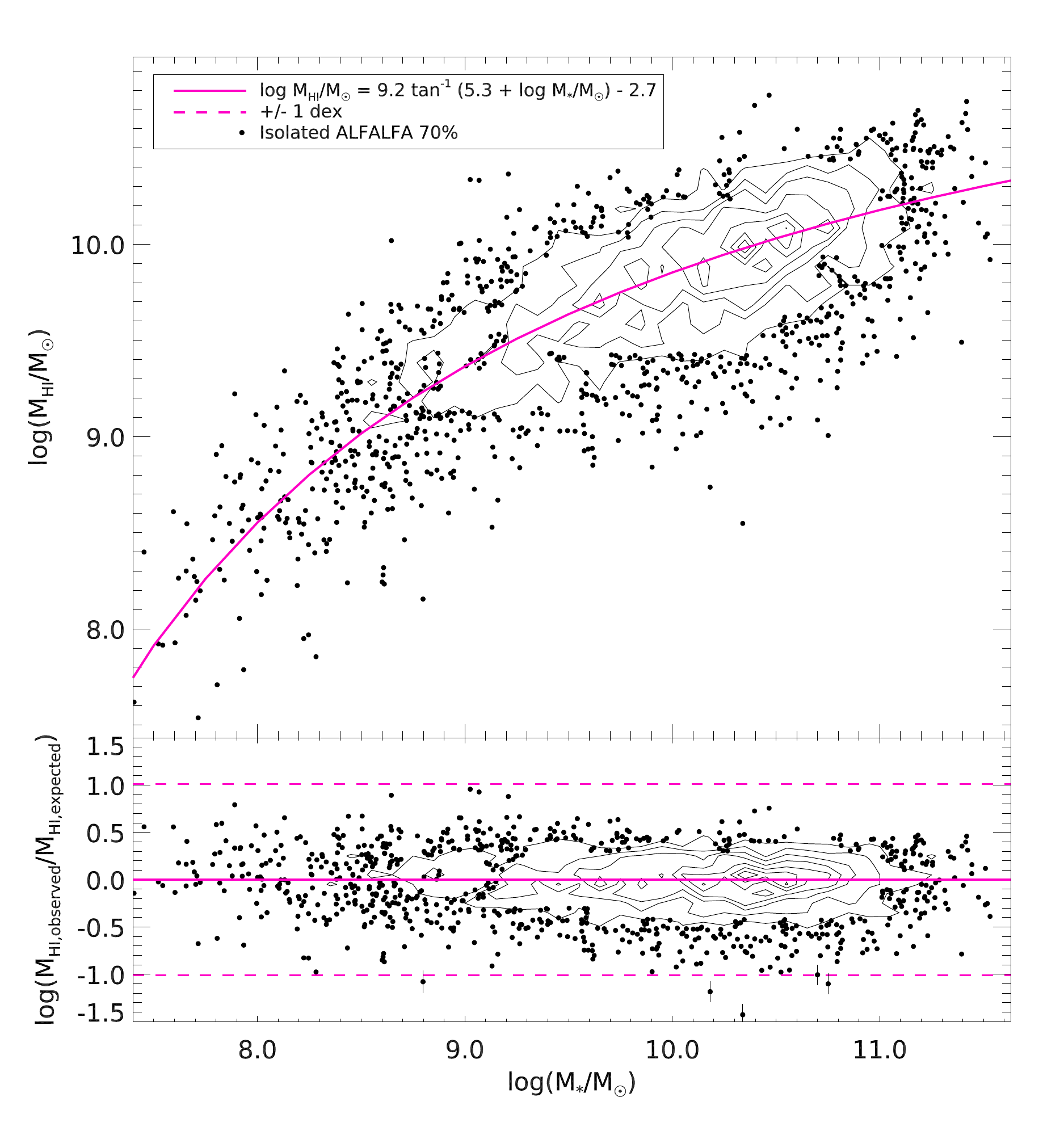}
\caption[ALFALFA 70\% Isolated \MHI\ to \MS\ Relation of Gas-Rich Galaxies]{Top: The isolated \MHI\ to \MS\ relation of gas-rich galaxies. Only isolated galaxies with detections from the 70\% ALFALFA survey are plotted here as black contours and dots. The fit from equation \ref{eq_m_star_m_hi} is plotted as the solid curve and noted in the legend. Bottom: \HI\ mass residual (\DMI~$< -1$~dex) as a function of stellar mass with the same symbols at the top panel. The $\pm 1$~dex (3$\sigma$) residuals are shown as dashed lines. }
\label{fig_m_star_m_hi}
\end{figure}

The observed cold gas content of galaxies is closely related to
their stellar mass, galaxy color, and stellar surface density
\citep[e.g.,][Paper I]{Catinella:2010eo, Huang:2012gk}. The baryon
content of isolated low-mass galaxies is often dominated by cold gas, likely
because galaxy formation is inefficient at low masses
\citep{Roychowdhury:2014jf}. A break in the relation between \MHI\ and
\MS\ occurs near log\,\MS~$= 9.0$, where the gas fraction increases to more
than 90\% at the lowest stellar masses. In Paper I, we fit a
broken power law to the $M_{\rm gas} = 1.4 M_{\rm \HI}$ to
\MS\ relation with a break at log\,\MS~$= 8.6$. In this work however, we
find that an arctangent function provides a better fit simultaneously
to the high and low-mass ends of the \MHI\ to \MS\ relation while
maintaining the break between high and low \MS\ regimes. This function
is not physically motivated but offers a better fit to the locus of
the distribution in \MHI\ at fixed \MS\ than a simple broken
power-law.

We fit the relation only to the isolated galaxies selected from the NSA catalog that have ALFALFA \HI\ data. The relation is measured with ALFALFA data only because of homogeneity and the survey's well-studied completeness and detection limits \citep[][\S\ 6]{Haynes:2011en}. We use only isolated galaxies 
to minimize the effects of environmental gas-depletion processes on this relationship.

To perform the fit, we bin the data in stellar mass and then fit to the weighted medians of the distribution of \MHI\ in each stellar mass bin. Because the distribution of \MHI\ as a function of stellar mass is log-normal with the tail of the distribution trending to smaller \MHI, this fitting method removes any spurious trends in the residuals in this relation, which is critical for our analysis below. We measure the following relation:

\begin{equation}
\log{\rm{M_{\HI}/M_{\odot}}} = \Azero~\rm{tan}^{-1}{(\Aone + \log{\rm{M_*/M_{\odot}}})} \Atwo.
\label{eq_m_star_m_hi}
\end{equation}

\noindent We calculate the residuals from the \MHI\ to \MS\ relation for both the ALFLFA sample and our \HI\ data as,

\begin{equation}
\Delta \rm{M_{\HI}} = \log{\rm{M_{HI, observed}/M_{HI, expected}}}.
\label{eq_m_star_m_hi_residual}
\end{equation}

\noindent We add \MHIResidErr~dex in quadrature to all \HI\ residual uncertainties and upper limits due to the uncertainty in the fit. 

We present the relation between \MHI\ and \MS\ in the top panel of
Figure \ref{fig_m_star_m_hi} for isolated galaxies with ALFALFA data
and the residuals (\DMI) between the observed \MHI\ and the predicted
\MHI\ in the bottom panel. Equation \ref{eq_m_star_m_hi} is plotted as
the solid curve. The dashed lines in the bottom panel are $\pm 1$~dex
from the relation, which roughly corresponds to the 3$\sigma$
scatter.   We use the $-1$~dex residual as a reference point throughout the paper as an arbitrary line of demarcation.


\subsection{Emission Line Measurements}
\label{subsec_agnhi_emldiag}

Optical emission line flux ratio diagnostics are often used to
differentiate between various physical phenomena that ionize the
interstellar medium of galaxies \citep{Baldwin:1981ev,
  Veilleux:1987dv}. Differentiating between excitation sources
(actively accreting black holes, star formation, shocks from AGB
stars, etc.) depends on the ionization mechanism
strength. \citet{Reines:2015dw} measure emission line strengths of \~67,000
galaxies selected from the NSA catalog. The authors cull the spectral
catalog with a S/N cutoff of 3 and an equivalent width cutoff of 1 for
the \Ha, \NII, \OIII\ emission lines and a S/N cutoff of 2 for the
\Hb\ line. They iteratively subtract continuum and absorption lines
using the models of \citet{Tremonti:2004ed} from the observed
spectrum, carefully masking out the emission lines. Next, a one- or
two-component Gaussian is fit to the \SII\ doublet and used as a
template for the remaining emission lines of interest, excluding
\OIII. We use the dataset from this study without any modifications.

\subsection{BPT emission line Diagnostics}
\label{subsec_agnhi_bpt_diag}

\begin{figure}[t!]
\epsscale{1.22}
\plotone{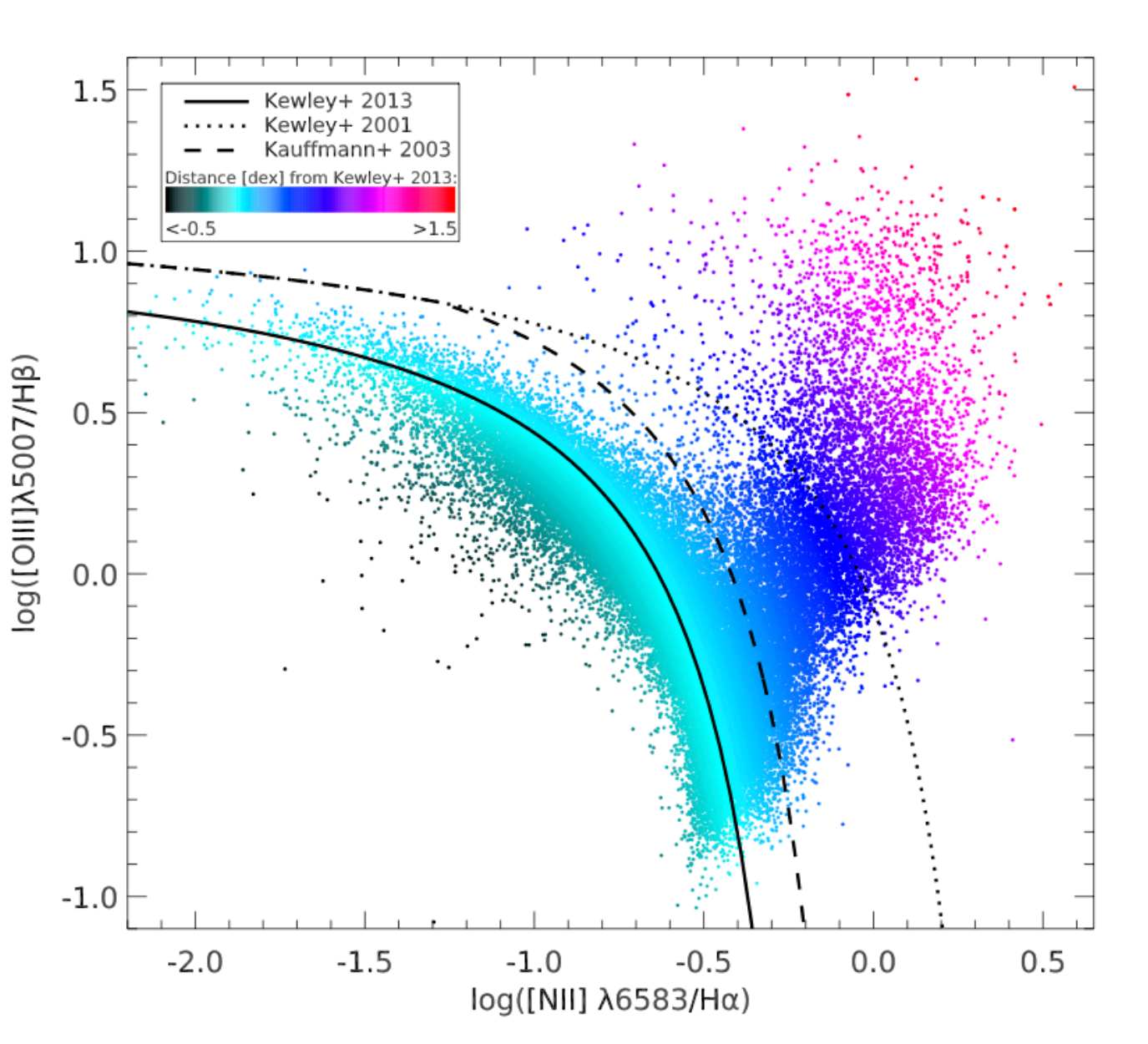}
\caption[\OIII/\Hb\ versus \NII/\Ha\ Emission Line BPT Diagram]{\OIII/\Hb\ versus \NII/\Ha\ emission line diagnostics BPT diagram with data from \citet{Reines:2015dw}, regardless of whether the galaxy has \HI\ data or not. The dotted line is the \citet{Kewley:2001cz} demarcation line, the dashed line is the \citet{Kauffmann:2003iz} demarcation line and the solid line marks the \citet{Kewley:2013ht} star-forming main sequence. Galaxies are color-coded by the perpendicular distance from the star-forming region as shown in the color bar.}
\label{fig_bpt_fit}
\end{figure}

We examine the emission line diagnostics of all galaxies from
\citet{Reines:2015dw}, regardless of environment or \HI\ data using
the \OIII/\Hb\ and \NII/\Ha\ narrow-line BPT diagram
\citep{Baldwin:1981ev, Veilleux:1987dv, Kewley:2006ib}. We adopt the
division into the star-forming, composite, and AGN categorization that
has been performed by \citet{Reines:2015dw} using the
\citet{Kewley:2001cz} and \cite{Kauffmann:2003iz} categorization. The
\citet{Kewley:2001cz} demarcation line represents the division between
a theoretical maximal starburst model and emission that can only be
explained by AGN activity. The \cite{Kauffmann:2003iz} demarcation
line is an empirical separation between the star forming sequence and
the AGN sequence. Galaxies below the \cite{Kauffmann:2003iz} line are
considered to be purely star forming and galaxies above the
\citet{Kewley:2001cz} line are considered Seyferts and LINERs. The
region between these two demarcation lines are defined as composite
spectra with both star formation and AGN emission.

Galaxies are commonly categorized using these divisions in BPT
space. This classification scheme has been calibrated with galaxies
more massive than the \citet{Reines:2013bp} sample.  For low mass
galaxies where AGN emission is weaker and may be more difficult to
disentangle from star formation emission, it is unclear where the
standard definitions are effective.   Lower mass galaxies are also, on average, more metal-poor which will affect the measured line ratios.   To further compound this problem,
the ratio of emission from star formation to AGN activity is a strong
function of the spatial resolution of the observations - where single
fiber observations can create aperture bias
\citep[e.g.,][]{Sharp:2010jl, Pracy:2014bm, Gomes:2016jm}. 
The mean fraction of the
surface area covered by the SDSS fiber compared to the 50\% Petrosian
radii is \highfifty\% and \lowfifty\% for isolated high and low-mass
galaxies, respectively.    We expect smaller, less
massive galaxies with AGN emission to shift towards the composite or
possibly the star-forming region of the BPT diagram. Thus
emission-line categories may significantly limit our analysis.

\begin{figure*}[t!]
\epsscale{1.2}
\plotone{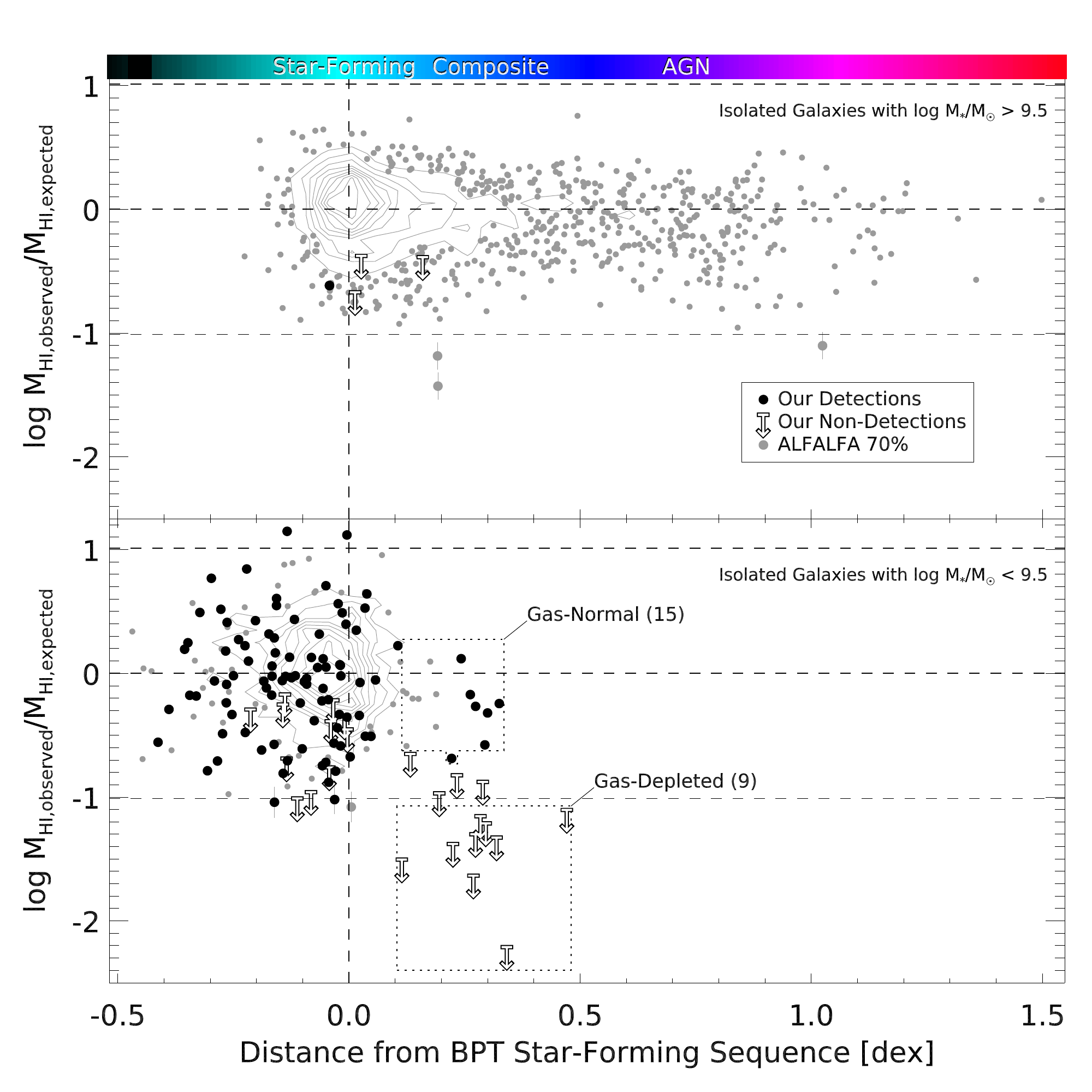}
\caption[BPT Star-forming Distance and \MHI\ Residuals for Isolated Galaxies]{Top: The residual (\DMI) in expected \MHI\ verses the observed \MHI\ as a function of the perpendicular distance from the BPT star-forming sequence defined in equation \ref{eq_kewley_three} for isolated, high-mass galaxies with $\LOGMS > 9.5$. Our observations are shown as black dots and our upper limits are shown as downward-pointing arrows. ALFALFA observations are shown as grey dots and contours. The color bar corresponds to the \DBPT\ as in Figure \ref{fig_bpt_fit}. The star-forming, composite, and AGN labels on the color bar represent the hard cutoff regions in the \OIII/\Hb\ versus \NII/\Ha\ emission line diagnostics BPT diagram as defined by \citet{Kauffmann:2003iz} and \cite{Kewley:2004dr}. Dashed horizontal and vertical lines mark the zeros of both axes. The additional horizontal lines at $\pm 1$~dex represent the $3 \sigma$ scatter in the fitted relation in \MHI\ versus \MS\ as in the bottom panel of Figure \ref{fig_m_star_m_hi}. Bottom: The same as the top panel but for isolated low-mass galaxies with $\LOGMS < 9.5$. In this stellar mass regime, we identify \nnint\ gas-depleted galaxies in the isolated sample with intermediate BPT distances. We also identify a sample of gas-normal galaxies that have been detected in \HI\ at similar \DBPT\ as the gas-depleted sample. We do not include the non-detections in the intermediate region in our analysis.}
\label{fig_bpt_dist_hi_resid}
\end{figure*}

%

It is therefore enlightening to examine galaxy properties relative to position on a BPT diagram as opposed to categorizations \citep[e.g.,][]{Kewley:2001cz, Kauffmann:2003iz, Kewley:2013ht, Zhu:2015dg}. For example, \cite{Kauffmann:2003iz} examine the \OIII/\Hb\ and \NII/\Ha\ narrow-line BPT space by defining a distance and an angle from a point where the star-forming sequence meets the plume of galaxies with AGN emission. They find that \OIII\ luminosity, a measure of AGN strength, does not increase with distance from this point, while \OIII\ luminosity decreases sharply at a characteristic angle. 

We measure the perpendicular distance from the star-forming sequence of the \OIII/\Hb\ and \NII/\Ha\ BPT diagram as defined by \citet{Kewley:2013ht} for each galaxy. The position of the \citet{Kewley:2013ht} fit is determined by local SDSS star-forming galaxies while the shape of the curve is based on data from their photoionization models (see their Figure 1),

\begin{equation}
\log{\left( \frac{[\rm O III]}{{\rm H}\beta} \right)} = \frac{0.61}{ \log(\rm{[NII]}/{\rm H}\alpha) +0.08 }+1.1.
\label{eq_kewley_three}
\end{equation}

\noindent For each galaxy, we define \DBPT\ as the minimum
perpendicular distance of each galaxy to equation
\ref{eq_kewley_three}. In Figure \ref{fig_bpt_fit}, we present the
\OIII]/\Hb\ versus \NII/\Ha\ emission line diagnostic BPT
  diagram. The data are color-coded by their distance from the
  star-forming sequence as defined above in equation
  \ref{eq_kewley_three}. This figure includes every galaxy with BPT
  data, regardless of whether or not there exists \HI\ data.

\subsection{Final Sample Description}
\label{subsec_galaxy_sample}

For the remainder of this work, we focus on a galaxy sample defined by the following criteria:
\begin{enumerate} 

\item  Galaxies selected from the NSA catalog passing
the quality cuts from Paper I.  The NSA includes galaxies in the redshift range $0.002 < z < 0.055$. 

\item  Galaxies are isolated given the criteria for low-mass
and high-mass galaxies given in \S\ \ref{subsec_agnhi_nsa}. 

\item Galaxies have \HI\ observations from Paper I, our \HI\ dataset or from
the ALFALFA 70\% catalog. 

\item Galaxies have well-measured emission line ratios from \citet{Reines:2015dw}. 

\end{enumerate}
The sample defined by combining the above criteria is not complete,
but does fully sample the BPT diagram. This final data set
consists of \nhiisobpt\ isolated galaxies with \HI\ and BPT data,
\alllmisowithhibpt\ of which we refer to as 'low-mass' $(\LOGMS < 9.5$).

\section{Results}
\label{sec_results}

\subsection{\HI\ Residuals as a Function of BPT Distance}
\label{subsec_agnhi_hibptresid}

In Figure \ref{fig_bpt_dist_hi_resid}, we present isolated galaxies
with both BPT and \HI\ data. We divide this sample at $\LOGMS = 9.5$,
as shown in the top and bottom panels. In this figure, we examine the
\MHI\ residual (\DMI) as a function of distance from the BPT
star-forming sequence (\DBPT). ALFALFA detections are shown as grey
points and contours. The \HI\ data from Paper I and our deeper
observations are shown as black dots for detections and downward
pointing arrows for upper limits. The color bar corresponds to
\DBPT\ values as in Figure \ref{fig_bpt_fit}. The star-forming,
composite, and AGN regions, as defined by \citet{Kewley:2001cz} and
\cite{Kauffmann:2003iz}, are labeled in the color bar at the median
positions of the \DBPT\ distributions of the entire sample for each of
the three categories.  These labels are only meant to act as a
guide. The $\pm 3\sigma$ scatter in the \HI\ residual, \DMI~$\pm
-1$~dex, is shown as horizontal dashed lines. Galaxies with
\DMI~$<-1$~dex are plotted with error bars. Given the conservative
ALFALFA upper limit estimates from \S\ \ref{subsubsec_agnhi_alf_hi},
we would not expect ALFALFA to detect a significant number of galaxies
below the \DMI~$= -1$~dex line.

The high-mass galaxy sample in the top panel of Figure \ref{fig_bpt_dist_hi_resid} shows no trend in \MHI\ as a function of \DBPT. While this is certainly an incomplete \HI\ sample due to the flux-limited nature of the ALFALFA survey, we can safely assume the incompleteness is the same for all galaxies across the entire BPT diagram because \DBPT\ is not dependent on heliocentric distance.  This lack of a trend supports previous studies (see \S\ \ref{sec_agnhi_intro}) suggesting that the global \HI\ content of high-mass galaxies is not significantly affected by AGN activity.

The low-mass galaxy sample is shown in the bottom panel of Figure \ref{fig_bpt_dist_hi_resid}.
This sample suffers from the same \HI\ incompleteness as the high-mass sample but, as in the high-mass case, we do not expect this incompleteness to be a function of \DBPT.    The peak in the distribution of BPT distances is offset by $-0.05$~dex from the locus of the star-forming sequence. This offset is likely due to the fact that low-mass galaxies are metal-poor compared to massive galaxies  (see \citet{Kewley:2013ht} equation 4).  Since the star-forming population of low mass galaxies is offset relative to the typical AGN demarcation lines, this further motivates our decision to examine galaxies are a function of BPT distance ($d_{\rm BPT}  $).

In the low-mass regime, we have followed-up many galaxies with deeper \HI\ data from Paper~I.  In the star-forming region, we have randomly sampled this population and do not observe any star-forming galaxies with \DMI~$< -1$~dex despite these deeper observations. Indeed, given the results from Paper I, we expect cold gas fractions greater than 20\% for all isolated galaxies with $7.0 < \LOGMS < 9.5$.  However our deeper \HI\ observations, do uncover a population of galaxies at intermediate BPT distances where $d_{\rm BPT} > $ \minbptin~dex and \HI\ upper limits \DMI~$<-1$~dex. These galaxies are ``gas-depleted" relative to galaxies in the same stellar mass range by an order of magnitude and they are inconsistent with the \HI\ distribution of the star-forming population. This gas-depleted galaxy sample with \DMI~$< -1$~dex may be indicative of an ionizing process other than star-formation that is depleting the ISM of \HI. 


In the same intermediate region of BPT distances where $d_{\rm BPT} > $ \minbptin~dex, we also find low-mass galaxies that are ``gas-normal". These gas-normal galaxies have \HI\ gas masses that are consistent both with the star-forming population and with galaxies in the same stellar mass range. This implies that whatever is causing these strong emission line ratios does not always significantly deplete the global \HI\ mass.



\subsection{Isolated Low-Mass Galaxies at Intermediate BPT Distances}
\label{subsec_agnhi_isointbpt}

\begin{figure*}[t!]
\epsscale{1.2}
\plotone{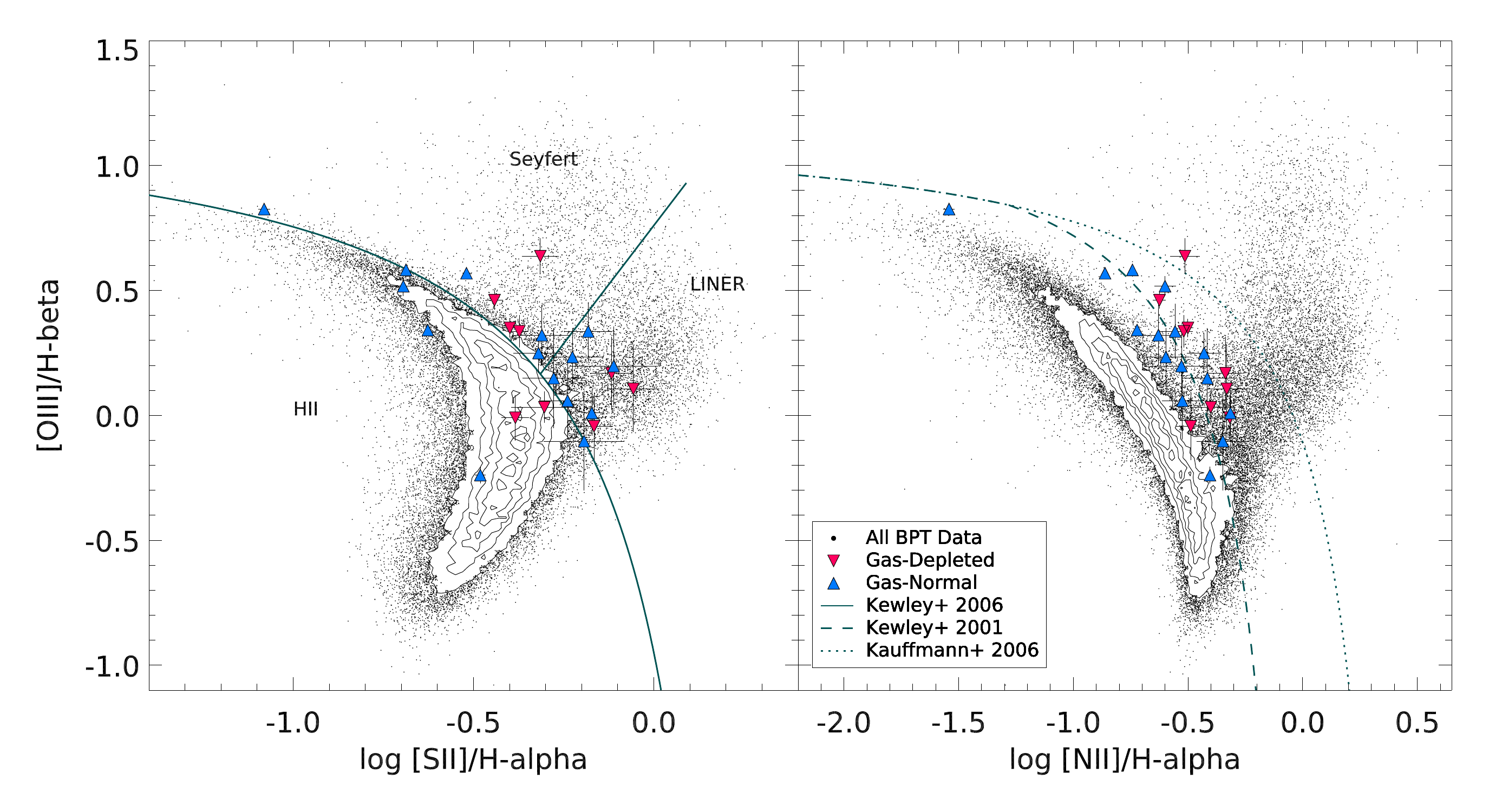}
\caption[\OIII/\Hb\ vs \SII/\Ha\ \& the \OIII/\Hb vs \NII/\Ha\ narrow-line diagrams]{Left: The \OIII/\Hb\ and \SII/\Ha\ narrow-line diagram. All galaxies with BPT data are plotted as black points and contours. The gas-depleted, isolated, low-mass sample is shown as red, downward-pointing triangles and the gas-normal, isolated, low-mass sample is shown as blue triangles (see \S\ \ref{subsec_agnhi_hibptresid}). The \citet{Kewley:2006ib} demarcation between Seyferts and LINERs as well as their separation from the HII region are shown as the green line and curve, respectively. The majority of gas-depleted galaxies are found above the \citet{Kewley:2006ib} demarcation line in both the Seyfert and the LINER regions. The gas-normal sample is similarly distributed, but trends towards the LINER region. The distribution of the two samples support the conclusion that the emission line ratios are due to ionization processes other than star formation. Right: The \OIII/\Hb vs \NII/\Ha\ BPT diagram. Data and symbols are the same as in the left panel. Curves are those defined as in Figure \ref{fig_bpt_fit}. Note that many of the gas-normal galaxies are located above the locus of the star-forming region but are categorized as ``star-forming" using the standard BPT classification scheme. 7 of the 9 gas-depleted galaxies are considered composites, 1 is star-forming and 1 is considered an AGN.}
\label{fig_bpt_sii}
\end{figure*}

 \input{tbl_int_samples.tex}

We find a set of gas-depleted, isolated, low-mass ($\LOGMS = 9.5$)
galaxies with signs of AGN activity.  We next compare the
gas-depleted galaxies at large \DBPT\ to the gas-normal galaxies in
the same \DBPT\ regime in order to explore the nature of the ionizing
power source. We also compare the properties of both the gas-depleted
sample and the gas-normal sample to a stellar-mass matched sample of
isolated galaxies.

In other to further understand the source of ionization in
our large \DBPT\ galaxies, we examine additional line diagnostics.
Active galaxies are often split into Seyfert and LINERs using the the
\OIII/\Hb\ and \SII/\Ha\ narrow-line BPT diagram
\citep{Heckman:1980wd}.   While Seyfert emission is associated with
accretion onto a black hole, LINER emission is more ambiguous, having
been detected outside the nucleus of many galaxies.  \citet{Kewley:2006ib}  identify a separation \SII/\Ha\ vs \OIII/\Hb\
BPT space between Seyfert and LINER emission.

In Figure \ref{fig_bpt_sii}, we present the \OIII/\Hb\ and \SII/\Ha\ narrow-line diagram in the left panel and revisit the \OIII/\Hb\ and \NII/\Ha\ BPT diagram in the right panel. All galaxies with emission line data are plotted as black points and contours. The gas-depleted, isolated, low-mass sample is shown as red, downward-pointing triangles and the gas-normal, isolated, low-mass sample is shown as blue triangles (see \S\ \ref{subsec_agnhi_hibptresid}). In the left panel, the \citet{Kewley:2006ib} demarcation curve separates Seyferts and LINERs from the HII region and the straight line splits the active galaxies into Seyferts and LINER branches, as labeled. 

In the left panel of Figure \ref{fig_bpt_sii}, the majority of the
gas-depleted and the gas-normal samples lie above and along the
extreme starburst curve of \citet{Kewley:2001cz, Kewley:2006ib}. Both
samples are therefore largely inconsistent with the star-forming galaxies. The gas-depleted sample above the curve is split between
the Seyfert and LINER regions, while the gas-normal sample is mostly
LINER-like. Figure \ref{fig_bpt_sii} provides further evidence that
the emission line measurements of the majority of the gas-depleted
sample and a fraction of the galaxies in the gas-normal sample are
partially due to ionization processes other than star formation (see
further discussion in \S\ \ref{sec_agnhi_discussion}). 



\begin{figure*}[t!]
\epsscale{0.92}
\plotone{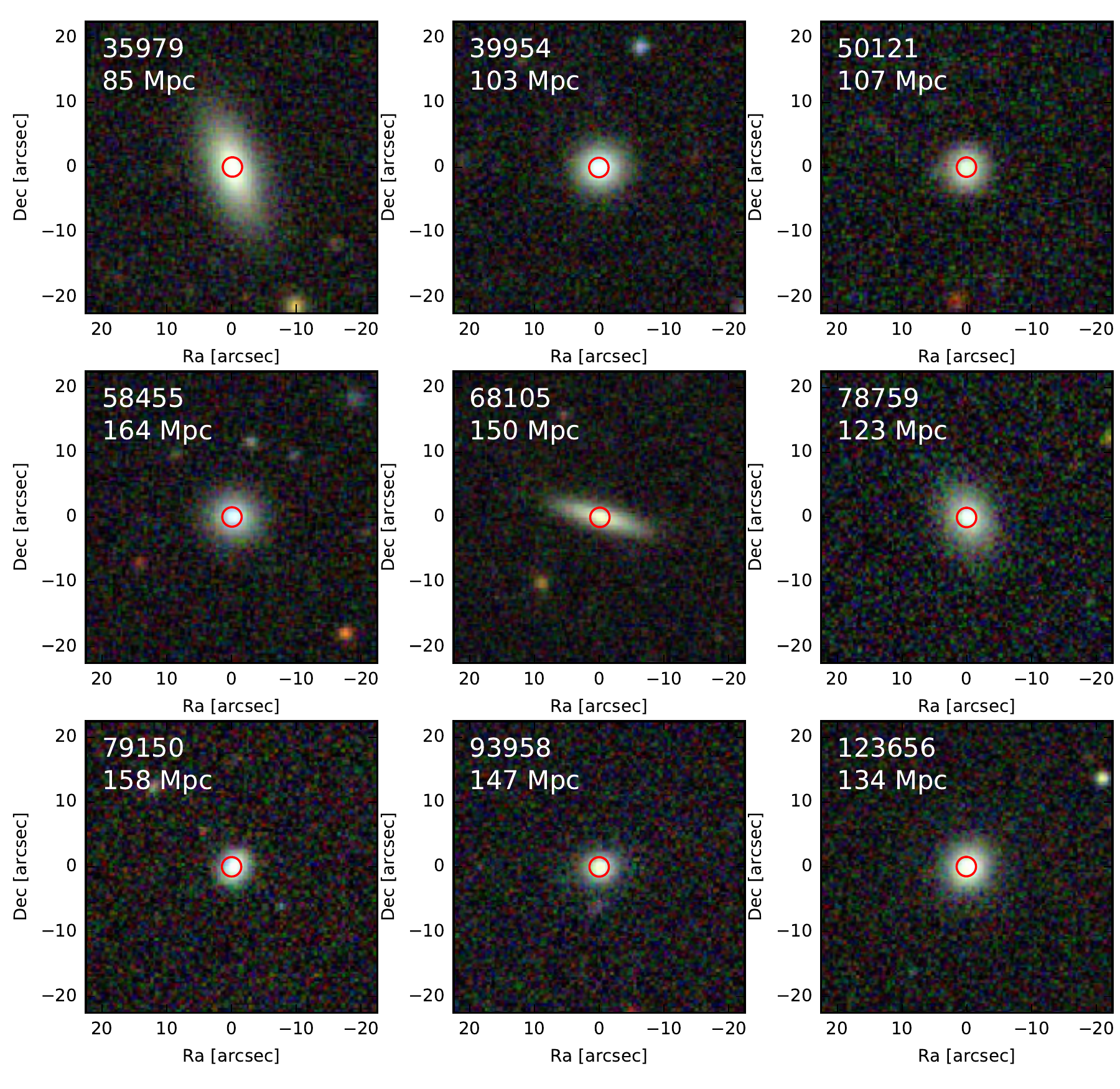}
\caption[Gas-depleted Isolated Sample Images]{SDSS images of gas-depleted, low-mass, isolated galaxies at large BPT distances. The NSAID is noted in the top right of each image. The SDSS fiber position and angular size on the galaxy is shown as a red circle. The heliocentric distance of each galaxy is noted below each NSAID. These gas-depleted galaxies are all red, compact and nucleated. 35979 and 68105 are red, but prolate. These galaxies are visually inconsistent with typical star-forming galaxies.}
\label{fig_int_stamps}
\end{figure*}

\begin{figure*}[t!]
\epsscale{0.95}
\plotone{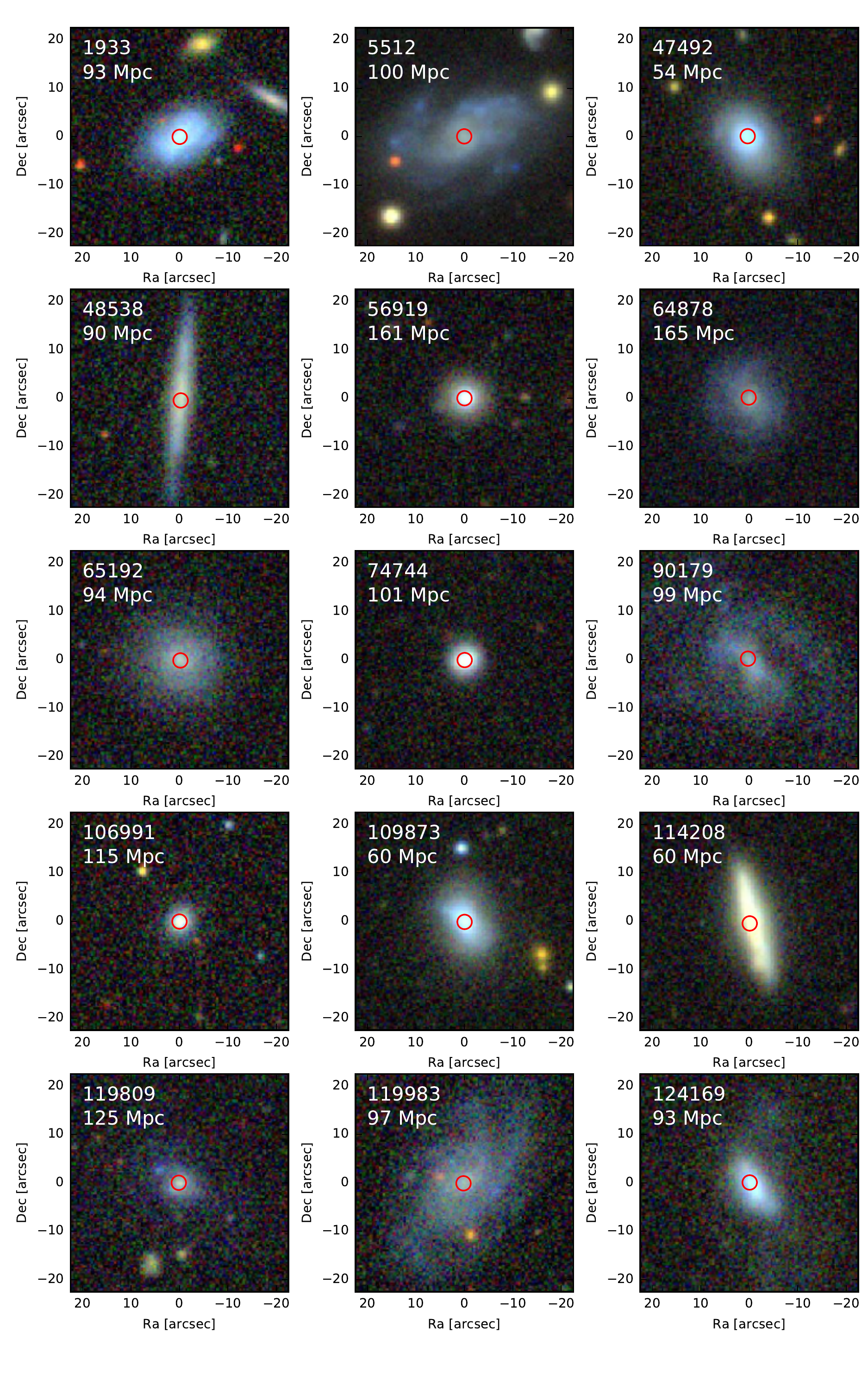}
\caption[Gas-normal Isolated Sample Images]{Same as in Figure \ref{fig_int_stamps} but for the gas-normal low-mass isolated galaxies at large BPT distances. This galaxy population is much more diverse than the gas-depleted sample. Many galaxies are blue and diffuse with some signs of structure and star formation and the SDSS fiber covers a small fraction of these galaxies. 56919, 74744, 106991 and 114208 are all visually similar to the gas-depleted sample.}
\label{fig_nor_stamps}
\end{figure*}

In order to further disentangle the source of the emission line ratios
and the relationship of these processes to the global \HI\ content of
these galaxies, we examine the galaxy properties of the two
samples. In Figure \ref{fig_int_stamps}, we present SDSS images of the
gas-depleted sample. The identifications used in each figure are the
NSAIDs from version 0.1.2 of the NSA catalog. The SDSS fiber positions
and sizes are plotted as red circles on each galaxy image. The
heliocentric distances are noted in each image. All of the
gas-depleted galaxies have similar visual properties (red, compact
and nucleated) aside from 35979 and 68105 which are red yet more
prolate in their respective orientations than the rest of the
gas-depleted sample. The SDSS fiber covers a significant fraction of
these galaxies' projected areas and these galaxies' spectra may
contain a significant amount of contamination from emission outside of
the galaxy centers (see \S\ \ref{subsec_agnhi_hibptresid}).

In Figure \ref{fig_nor_stamps}, we present images of the gas-normal
sample. This sample is diverse in galaxy properties relative to the
gas-depleted sample. Objects 56919, 74744, 106991 and 114208 are similar in
appearance to the gas-depleted population and the SDSS fiber again
covers a significant fraction of these galaxies. The remaining
galaxies tend to be blue and more diffuse with some signs of spiral
structure and larger sizes than the SDSS fiber. For many of these
galaxies there are visual signs of both diffuse and concentrated star
formation regions.

\begin{figure*}[t!]
\epsscale{1.18}
\plotone{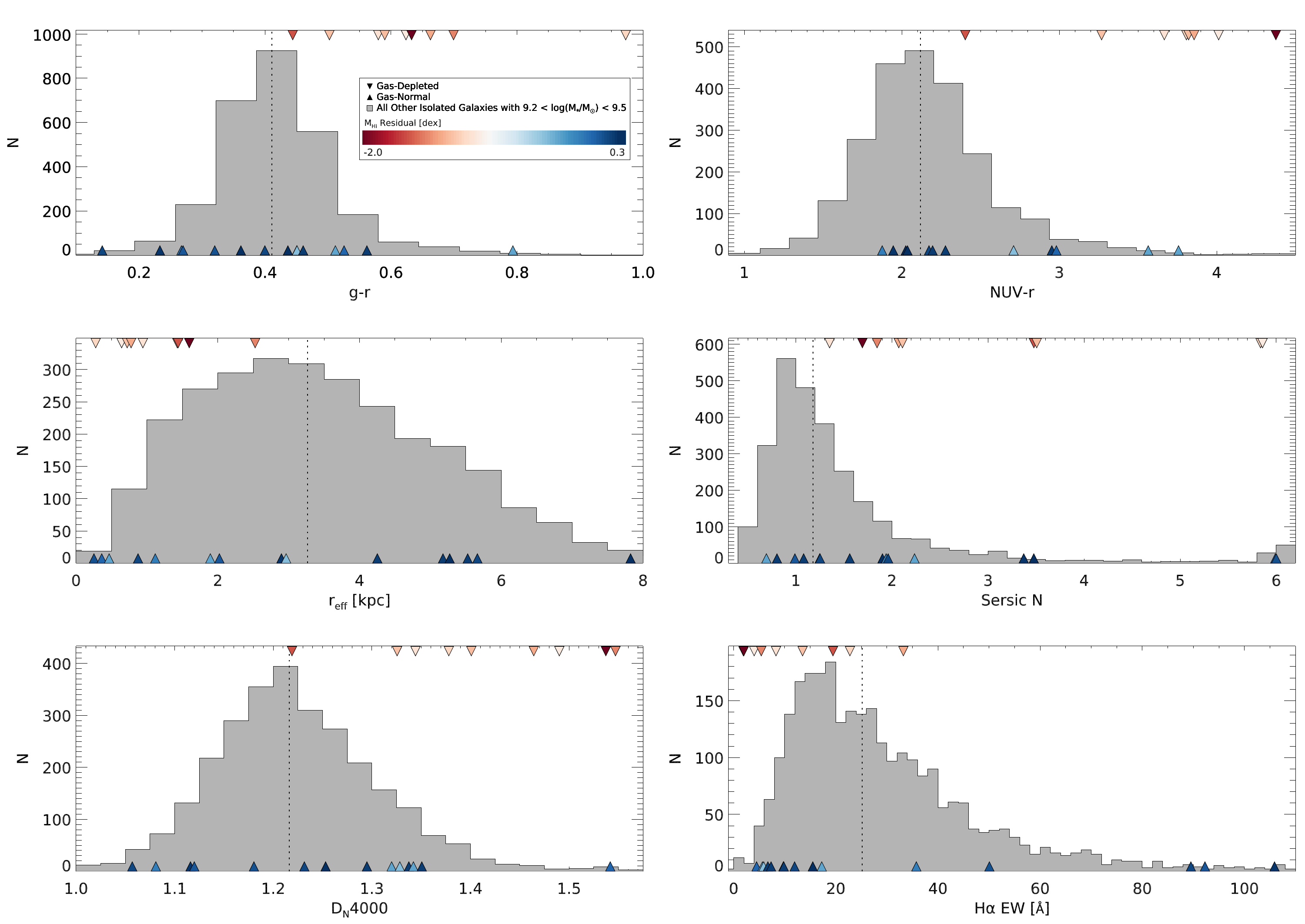}
\caption[Galaxy Property Distributions]{Distribution of gas-depleted and gas-normal isolated galaxies relative to the distribution of all isolated galaxies in the cleaned NSA catalog within the stellar mass range $9.2 < \LOGMS < 9.5$. The median of each distribution is shown as a vertical dashed line. Gas-depleted, isolated, low-mass galaxies are identified as downward pointing triangles. Gas-normal, isolated, low-mass galaxies are identified as upward pointing triangles. Both sets of galaxies are color-coded according to their \HI\ mass residual as noted in the color bar. The grey histogram is that of all isolated NSA galaxies within the stellar mass range of the isolated low-mass gas-depleted sample regardless of whether BPT or \HI\ data exist and does not include the gas-depleted or the gas-normal samples. From the top left panel, moving left to right and downward are the g-r colors, NUV-r colors, effective radii, Sersi\'c index, D$_{n}4000$ and \Ha\ equivalent widths. Note that we also examine \OIII\ luminosity (not shown) and both samples are consistent with the general distribution.}
\label{fig_int_hist}
\end{figure*}

In Figure \ref{fig_int_hist}, we examine the distributions of g-r colors, NUV-r colors, effective radius, Sersi\'c index, D$_{n}4000$, and H$\alpha$ equivalent width of both the gas-normal and the gas-depleted low-mass samples relative to all isolated galaxies in the NSA catalog within the same stellar mass range (with $9.2 < \LOGMS < 9.5$). The gas-normal sample is shown as upward pointing triangles, color-coded by \DMI. The gas-depleted sample is shown as downward pointing triangles, also color-coded by \DMI. Isolated galaxies from the NSA catalog, excluding the two samples of interest, are shown as grey histograms in each panel regardless of whether BPT or \HI\ data exists. The median of each distribution is shown as a vertical dashed line. We present these and other relevant properties of the gas-depleted and gas-normal galaxy samples in Table \ref{tbl_int_samples}.

Confirming our visual analysis, the distribution of the gas-depleted
sample properties is inconsistent with being drawn from the overall
star-forming population.  For each property shown in
Figure~\ref{fig_int_hist}, we test the null hypothesis that the
gas-depleted/gas-normal sample and the full isolated sample of
galaxies in the same stellar mass range are drawn from the same
underlying population.  We use a variation of the Student's t-test
without the assumption that the variances of the two populations are
equal.  The gas-depleted galaxies are not drawn from the same
distribution as the overall population with greater 95\% confidence for
all 6 properties shown in Figure 7.  For the gas-normal population,
the distributions are consistent (that is, we cannot reject the null
hypothesis that they are drawn from different underlying
populations).


We also examine \OIII\ luminosity in the same fashion as above (not
shown). Both the gas-normal and gas-depleted samples are consistent
with the general distribution of \OIII\ luminosity. However, the
general distribution become restricted to galaxies that have emission
line measurements from \citet{Reines:2015dw}.  \cite{Kauffmann:2003iz} also do not find that \OIII\ luminosity changes as a function of a galaxy's distance from locus of the star-forming BPT region.

\section{Discussion}
\label{sec_agnhi_discussion}

We find a set of galaxies with $9.2 < \LOGMS < 9.5$ that are both
isolated and gas-poor. These galaxies are located primarily in the
composite region of the \OIII/\Hb\ versus \NII/\Ha\ BPT diagram and
the LINER/Seyfert regions of the \OIII/\Hb\ and \SII/\Ha\ BPT
diagram. These gas-depleted galaxies are also compact and red with
older stellar populations and no obvious signs of star formation.  

This result is surprising because isolated galaxies in this mass range
are
generally gas-rich, diffuse, blue and show ongoing star
formation. Indeed, \citet{Geha:2012eu} find that less than 2\% of
isolated galaxies with $9.2 < \LOGMS < 9.5$  are quiescent. If
these galaxies are truly not forming stars, they are extremely rare
cases of quenched, low-mass galaxies in isolation. Here we explore the
possible physical processes behind the emission line ratios and what
may have removed the gas from these galaxies.   We note that these
galaxies are defined as at least 1.5\,Mpc from any massive galaxy, such that 
 the usual environmental processes which dominate in group and cluster
 environments are unlikely to be active.

A straightforward interpretation is that these galaxies are depleted of \HI\ because of AGN feedback. In this scenario, the line ratios that we observe are excited by accretion onto the central black hole, and energy from accretion has successfully coupled to the gas in the galaxy on large scales and removed and/or destroyed the gas in the galaxy. Accretion energy is observed to drive large scale outflows in molecular \citep[e.g.,][]{Alatalo:2011go, Veilleux:2013du}, warm ionized \citep[e.g.,][]{Liu:2013hs, Harrison:2012kf}, and hot (warm absorbers, BALs) gas \citep[e.g.,][]{Crenshaw:2003hz, Reeves:2009ep}. The energy may couple to the gas via radiation pressure on dust \citep[e.g.,][]{Thompson:2005dq}, via thermal heating \citep[e.g.,][]{Weinberger:2016gx} or via input of kinetic energy \citep[e.g.,][]{Choi:2012ft}.
 
Many of the gas-normal and gas-depleted galaxies are classified as LINERs. While LINER emission in massive galaxies has been directly linked with black hole accretion, there are many instances of off-nuclear LINERs (see \citet{Ho:2008kz} and references therein). Even in cases with unambiguous signs of accretion, it is not clear that the AGN provide sufficient photons to photoionize the gas \citep{Eracleous:2010jv, Yan:2012jn}. A number of photoionization sources may explain the observed line ratios. Shocks and post-AGB (pAGB) stars are possible sources of excitation for the gas. In the case of pAGB stars, we would expect \Ha\ equivalent widths less than 3\AA\ if pAGB stars are causing a diffuse stellar emission \citep{Binette:1994tc, Belfiore:2016dva, Belfiore:2015kh}. The \Ha\ equivalent width distribution suggests that the majority of the line emission cannot be excited by pAGB stars, leaving shocks or AGN as the most viable options.

If AGN activity is not responsible for heating or removing the cold
gas, perhaps the compact structure of the gas-depleted galaxies
provides a clue to their nature. Star formation may have been
extremely bursty and destroyed or otherwise consumed the cold
gas. Compact galaxies can indeed experience extreme outflows caused by
compact stellar feedback \citep{DiamondStanic:2012cb}. If most of the
star-forming cold gas has been removed, we may be seeing the remaining
shocked gas from the outflows. This scenario could explain both the
gas depletion and the line ratios without invoking an AGN. The three
compact, gas-normal galaxies at large BPT distances will be
interesting to study for signs of higher star-formation rates or
larger outflow velocities than their more diffuse
counterparts. Finally,  higher spatial resolution imaging and spectroscopy of both the star-forming and
gas-depleted compact galaxies would be enlightening, to determine
whether they have a diffuse disk component or central point source.
	
\section{Summary}
\label{sec_agnhi_summary}

In this work, we study the \HI\ masses of galaxies with signs of AGN activity as measured by optical emission lines. We compile a catalog of isolated galaxies with optical emission lines measured from SDSS DR8 spectra and single-dish \HI\ observations both from the ALFALFA 70\% data release and from Paper I. We obtain new, deeper \HI\ observations with a focus on isolated, low-mass galaxies selected from the \citet{Reines:2013bp} ``dwarf" galaxy sample with signs of AGN activity. We measure a new \MS-\MHI\ relation using ALFALFA 70\% data as in Equation \ref{eq_m_star_m_hi}. We also measure the distance of each galaxy with \OIII/\Hb\ and \NII/\Ha\ data from the Baldwinâ Phillipsâ Terlevich (BPT) star-forming sequence (\DBPT). We use this distance as a measurement of AGN activity strength and compare \MHI\ as a function of \DBPT.  Our results are summarized as follows: 

\begin{enumerate}
\item{We identify a set of  isolated, low-mass galaxies with $\LOGMS < 9.5$ and $d_{\rm BPT} > $ \minbptin~dex that is significantly gas-depleted.}
\item{7 of the 9 gas-depleted galaxies are located above the \HII\ demarcation line of \citet{Kewley:2006ib} on the \OIII/\Hb\ and \SII/\Ha\ diagram, 3 of which are categorized as LINERs and 4 as Seyferts.}
\item{The gas-depleted, low-mass, isolated galaxies with large \DBPT\ are red and compact with old stellar populations and no distinguishable signs of star formation. The property distributions of the gas-depleted sample are inconsistent with the general population of isolated galaxies in the same stellar mass range.}
\item{We compare the gas-depleted sample to a population of 15 gas-normal, low-mass, isolated galaxies with large \DBPT. This sample consists primarily of star-forming, late-type galaxies. The majority of these galaxies are consistent with the general population of isolated galaxies in the same stellar mass range. However, we find 3 gas-normal galaxies with similar visual and structural properties as the gas-depleted population. We find a similar fraction of these gas-normal galaxies in the LINER and Seyfert regions on the \OIII/\Hb\ and \SII/\Ha\ BPT diagram as the gas-depleted sample.}
\end{enumerate}

Because these galaxies are very isolated (more distance than 1.5\,Mpc
from a more massive neighbor),  environmental process cannot explain
the emission line ratios or \HI\ masses in the gas-depleted sample.
We suggest two  possibilities below.

In the first case, the energy from black hole accretion successfully couples to the cold gas and causes both the emission lines and the depletion of \HI. The combination of deep radio and X-ray data would be useful to test the nature of these sources. As far as we are aware, no observations in radio continuum or X-ray exist for these galaxies.   

In the second case, the compact structural properties of the
gas-depleted sample may indicate a much more efficient and energetic
star formation history than other galaxies in the same stellar mass
regime. It could be that the shocks from this star formation generated
the emission lines and removed the cold gas from these
galaxies. Deeper imaging and resolved spectroscopy may help examine
the nature of these compact galaxies.  Star formation histories for both the gas-normal and gas-depleted compact galaxies would also help us determine the nature of these galaxies.

\begin{acknowledgements}

We thank Frank van den Bosch and Pieter van Dokkum for comments on this work.
J.D.B.~acknowledges support from the Gruber
Foundation and the National Science
Foundation Graduate Research Fellowship Program under Grant
No.~DGE-1122492.   
The Arecibo Observatory is operated by SRI International under a
cooperative agreement with the National Science Foundation
(AST-1100968), and in alliance with Ana G. M\'{e}ndez-Universidad
Metropolitana, and the Universities Space Research Association. The
National Radio Astronomy Observatory is a facility of the National
Science Foundation operated under cooperative agreement by Associated
Universities, Inc. Funding for SDSS-III has been provided by the
Alfred P. Sloan Foundation, the Participating Institutions, the
National Science Foundation, and the U.S. Department of Energy Office
of Science. The SDSS-III web site is http://www.sdss3.org/. 

\end{acknowledgements}

\acknowledgements


\bibliographystyle{apj}
\bibliography{agnhi_bibliography}

\end{document}

%% file: m_star_m_hi_fit.tex
\def\Azero{9.2}
\def\Aone{5.3}
\def\Atwo{-2.7}
\def\MHIResidErr{0.1}

%% file: catalog_stats.tex
\def\nnsaisoall{36708}
\def\nnsaisolow{8074}
\def\nnsaisohigh{28621}
\def\nalfalfarms{2.3}
\def\npostpaperone{30}
\def\maxinttime{60}
\def\mininttime{5}
\def\meaninttime{16}
\def\alllmisowithhibpt{867}
\def\nnint{9}
\def\minbptin{0.11}
\def\nnondetectall{65}
\def\nhiisobpt{2855}

%% file: fiber.tex
\def\highfifty{22}
\def\lowfifty{32}

%% file: tbl_new_hi_obs.tex
\begin{turnpage}
\begin{deluxetable*}{lcccccccccccc}
\tabletypesize{\tiny}
\tablecaption{New Arecibo HI Observations Having Data from \citet{Reines:2015dw}}
\tablewidth{0pt}
\tablehead{
\colhead{NSAID} &
\colhead{$\alpha$} &
\colhead{$\delta$} &
\colhead{$z$} &
\colhead{$S_{21}$} &
\colhead{$\sigma_{\rm{S_{21}}}$} &
\colhead{$W_{20}$} &
\colhead{$\LOGMHI$} &
\colhead{$\LOGMS$} &
\colhead{$D$} &
\colhead{$t_{\rm{int}}$} &
\colhead{$d_{\rm host}$} &
\colhead{BPT Classification}\\
\colhead{} &
\colhead{(h:m:s)} &
\colhead{($^\circ$:':'')} &
\colhead{} &
\colhead{(mJy~km/s)} &
\colhead{(mJy)} &
\colhead{(km/s)} &
\colhead{} &
\colhead{} &
\colhead{(Mpc)} &
\colhead{(s)} &
\colhead{(Mpc)} &
\colhead{}\\
\colhead{(1)} &
\colhead{(2)} &
\colhead{(3)} &
\colhead{(4)} &
\colhead{(5)} &
\colhead{(6)} &
\colhead{(7)} &
\colhead{(8)} &
\colhead{(9)} &
\colhead{(10)} &
\colhead{(11)} &
\colhead{(12)} &
\colhead{(13)}}
\startdata
1933 & 13:38:15.39 & -00:23:54.79 & 0.0220 & 736.5 $\pm$ 23.8 & 0.9 & 178 $\pm$ 6&9.2 $\pm$ 7.9 & 9.1 & 93 & 600 & 1.5 & Comp\\
13496 & 10:54:47.89 & +02:56:52.51 & 0.0222 & 890.7 $\pm$ 31.8 & 0.8 & 165 $\pm$ 10&9.3 $\pm$ 8.0 & 8.9 & 94 & 900 & 2.3 & SF~(broad~\Ha)\\
31601 & 07:37:43.65 & +24:09:34.45 & 0.0249 & 694.0 $\pm$ 25.1 & 1.3 & 142 $\pm$ 8&9.3 $\pm$ 8.0 & 8.5 & 108 & 300 & 1.6 & SF\\
39954 & 09:47:05.72 & +05:01:59.91 & 0.0242 & $<$49.4 & 0.8 & -&$<$8.1 & 9.2 & 103 & 1800 & 2.1 & Comp\\
50121 & 08:15:49.08 & +25:47:01.49 & 0.0248 & $<$37.7 & 0.7 & -&$<$8.0 & 9.2 & 107 & 2700 & 1.9 & Comp\\
56919 & 15:30:56.86 & +31:36:49.49 & 0.0365 & 371.4 $\pm$ 25.2 & 0.6 & 213 $\pm$ 29&9.4 $\pm$ 8.2 & 9.4 & 161 & 1200 & 3.2 & Comp\\
57536 & 15:47:03.21 & +34:13:19.48 & 0.0395 & $<$59.9 & 0.8 & -&$<$8.6 & 9.5 & 175 & 2100 & 7.1 & Comp\\
58455 & 16:32:10.87 & +29:01:30.60 & 0.0370 & $<$11.2 & 1.0 & -&$<$7.9 & 9.4 & 164 & 1080 & 2.9 & Comp\\
64878 & 09:27:47.64 & +33:12:33.17 & 0.0375 & 394.8 $\pm$ 15.2 & 0.9 & 74 $\pm$ 3&9.4 $\pm$ 8.1 & 9.4 & 165 & 300 & 3.1 & Comp\\
65192 & 09:50:20.09 & +36:04:46.56 & 0.0218 & 340.4 $\pm$ 12.4 & 1.1 & 68 $\pm$ 2&8.9 $\pm$ 7.6 & 9.3 & 94 & 600 & 1.7 & Comp\\
68105 & 15:30:27.30 & +28:31:20.71 & 0.0341 & $<$19.6 & 1.8 & -&$<$8.0 & 9.5 & 150 & 600 & 1.7 & SF\\
71340 & 14:29:02.53 & +10:23:59.23 & 0.0298 & $<$181.7 & 2.5 & -&$<$8.9 & 9.4 & 129 & 300 & 1.6 & SF\\
74744 & 10:55:55.13 & +13:56:16.97 & 0.0238 & 346.0 $\pm$ 16.1 & 0.9 & 65 $\pm$ 5&8.9 $\pm$ 7.7 & 9.2 & 101 & 600 & 1.8 & Comp\\
77431 & 13:04:34.92 & +07:55:05.10 & 0.0480 & $<$132.4 & 1.9 & -&$<$9.1 & 9.0 & 211 & 300 & 0.1 & AGN\\
79150 & 14:55:18.51 & +08:16:06.42 & 0.0361 & $<$34.8 & 0.7 & -&$<$8.3 & 9.4 & 158 & 2400 & 1.5 & Comp\\
84550 & 08:57:45.12 & +25:23:47.21 & 0.0133 & 826.2 $\pm$ 25.7 & 1.7 & 78 $\pm$ 4&8.9 $\pm$ 7.8 & 9.2 & 62 & 300 & 3.6 & SF\\
90179 & 10:47:58.08 & +33:55:37.13 & 0.0228 & 2114.2 $\pm$ 23.2 & 1.4 & 150 $\pm$ 2&9.7 $\pm$ 8.3 & 9.4 & 99 & 300 & 2.4 & Comp\\
91579 & 12:03:25.68 & +33:08:46.16 & 0.0349 & 138.6 $\pm$ 11.7 & 0.3 & 129 $\pm$ 17&8.9 $\pm$ 7.8 & 9.0 & 153 & 2400 & 1.6 & SF~(broad~\Ha)\\
93582 & 13:28:14.70 & +29:57:09.66 & 0.0095 & 238.9 $\pm$ 15.2 & 0.8 & 73 $\pm$ 5&8.0 $\pm$ 7.0 & 7.7 & 40 & 600 & 2.2 & SF\\
93958 & 11:31:29.21 & +35:09:59.01 & 0.0337 & $<$32.9 & 0.8 & -&$<$8.2 & 9.3 & 147 & 2400 & 1.5 & Comp\\
101949 & 11:43:02.42 & +26:08:18.99 & 0.0230 & $<$51.1 & 0.7 & -&$<$8.1 & 9.3 & 99 & 2400 & 1.0 & AGN\\
104334 & 13:12:19.26 & +27:07:37.26 & 0.0149 & 609.1 $\pm$ 43.4 & 1.3 & 141 $\pm$ 29&8.8 $\pm$ 7.8 & 8.5 & 68 & 300 & 2.4 & SF\\
104527 & 13:32:45.62 & +26:34:49.35 & 0.0470 & $<$63.5 & 0.9 & -&$<$8.8 & 9.4 & 208 & 1200 & 1.1 & AGN\\
105376 & 08:40:25.54 & +18:18:59.04 & 0.0150 & $<$99.1 & 1.6 & -&$<$8.1 & 9.3 & 69 & 600 & 1.2 & AGN\\
106991 & 10:04:23.34 & +23:13:23.37 & 0.0266 & 453.1 $\pm$ 15.5 & 0.5 & 176 $\pm$ 10&9.2 $\pm$ 7.9 & 9.2 & 115 & 1500 & 3.2 & Comp\\
107272 & 10:09:35.67 & +26:56:49.02 & 0.0144 & 154.0 $\pm$ 24.0 & 0.5 & 99 $\pm$ 36&8.2 $\pm$ 7.4 & 8.5 & 66 & 1200 & 1.2 & AGN\\
109397 & 07:58:12.44 & +11:01:14.18 & 0.0078 & 7217.1 $\pm$ 65.4 & 1.9 & 109 $\pm$ 2&9.4 $\pm$ 8.6 & 9.1 & 38 & 300 & 6.4 & SF\\
111644 & 11:05:03.97 & +22:41:23.43 & 0.0246 & $<$127.2 & 2.2 & -&$<$8.5 & 9.1 & 106 & 300 & 0.6 & AGN\\
118505 & 12:18:13.44 & +20:04:36.82 & 0.0457 & $<$31.7 & 0.8 & -&$<$8.5 & 9.3 & 201 & 1800 & 1.9 & Comp\\
123656 & 14:20:44.94 & +22:42:36.95 & 0.0307 & $<$51.1 & 0.8 & -&$<$8.3 & 9.3 & 134 & 1800 & 1.6 & AGN
\enddata
\tablecomments{Column definitions are: (1) NSAID, (2) right ascension,
  (3) declination, (4) redshift, (5) integrated 21-cm flux density,
  (6) rms per spectral element of the 21-cm flux after smoothing, (7)
  20\% HI line width, (8) HI mass, (9) stellar mass, (10) heliocentric
  distance, (11) total on-source integration time of 21-cm observation, (12) 2D projected distance from this galaxy to a more massive galaxy with $\LOGMS > 10$ (see \S\ \ref{subsec_agnhi_nsa}), (13) BPT category determined in \citet{Reines:2015dw} and \citet{Reines:2013bp},}
\label{tbl_new_hi_obs}
\end{deluxetable*}
\end{turnpage}


%% file: tbl_int_samples.tex
\begin{turnpage}
\begin{deluxetable*}{lccclcccccccccc}
\tabletypesize{\tiny}
\tablecaption{Gas-Depleted and Gas-Normal Isolated, Low-Mass Galaxy Samples}
\tablewidth{0pt}
\tablehead{
\colhead{NSAID} &
\colhead{$\alpha$} &
\colhead{$\delta$} &
\colhead{$z$} &
\colhead{Type} &
\colhead{$d_{\rm{BPT}}$} &
\colhead{$\Delta M_{\rm{HI}}$} &
\colhead{$\LOGMS$} &
\colhead{D} &
\colhead{g-r} &
\colhead{NUV-r} &
\colhead{$r_{\rm{eff}}$} &
\colhead{Sersi\'c N} &
\colhead{$D_{\rm{N}}4000$} &
\colhead{\Ha\ EW}\\
\colhead{} &
\colhead{(h:m:s)} &
\colhead{($^\circ$:':'')} &
\colhead{} &
\colhead{} &
\colhead{dex} &
\colhead{dex} &
\colhead{} &
\colhead{Mpc} &
\colhead{} &
\colhead{} &
\colhead{kpc} &
\colhead{} &
\colhead{} &
\colhead{\AA}\\
\colhead{(1)} &
\colhead{(2)} &
\colhead{(3)} &
\colhead{(4)} &
\colhead{(5)} &
\colhead{(6)} &
\colhead{(7)} &
\colhead{(8)} &
\colhead{(9)} &
\colhead{(10)} &
\colhead{(11)} &
\colhead{(12)} &
\colhead{(13)} &
\colhead{(14)} &
\colhead{(15)}}
\startdata
\multicolumn{15}{c}{Gas-Depleted}\\
35979 & 08:20:13.93 & +30:25:03.09 & 0.0196 & Comp & 0.34 $\pm$ 0.09 & $<$-2.2 & 9.5 & 85 & 0.63$\pm$0.03 & 4.38$\pm$0.08 & 1.60$\pm$0.03 & 1.7 & 1.54$\pm$0.03 & 2.0$\pm$0.3\\
39954 & 09:47:05.72 & +05:01:59.91 & 0.0242 & Comp & 0.27 $\pm$ 0.04 & $<$-1.3 & 9.2 & 103 & 0.50$\pm$0.03 & 3.27$\pm$0.06 & 0.72$\pm$0.01 & 2.1 & 1.33$\pm$0.03 & 13.5$\pm$0.5\\
50121 & 08:15:49.08 & +25:47:01.49 & 0.0248 & Comp & 0.23 $\pm$ 0.09 & $<$-1.4 & 9.2 & 107 & 0.66$\pm$0.03 & 3.86$\pm$0.09 & 0.78$\pm$0.02 & 2.1 & 1.46$\pm$0.05 & 33.2$\pm$0.9\\
58455 & 16:32:10.87 & +29:01:30.60 & 0.0370 & Comp & 0.27 $\pm$ 0.03 & $<$-1.6 & 9.4 & 164 & 0.44$\pm$0.03 & 2.40$\pm$0.05 & 1.44$\pm$0.02 & 3.5 & 1.22$\pm$0.02 & 19.4$\pm$0.7\\
68105 & 15:30:27.30 & +28:31:20.71 & 0.0341 & SF & 0.11 $\pm$ 0.07 & $<$-1.5 & 9.5 & 150 & 0.70$\pm$0.03 & - & 2.53$\pm$0.04 & 1.8 & 1.55$\pm$0.07 & 5.4$\pm$0.5\\
78759 & 14:12:43.72 & +08:22:16.06 & 0.0286 & Comp & 0.32 $\pm$ 0.10 & $<$-1.3 & 9.4 & 123 & 0.59$\pm$0.03 & 3.82$\pm$0.20 & 1.43$\pm$0.02 & 3.5 & 1.40$\pm$0.02 & 1.9$\pm$0.3\\
79150 & 14:55:18.51 & +08:16:06.42 & 0.0361 & Comp & 0.28 $\pm$ 0.04 & $<$-1.2 & 9.4 & 158 & 0.58$\pm$0.03 & 3.67$\pm$0.31 & 0.94$\pm$0.01 & 1.4 & 1.34$\pm$0.03 & 8.3$\pm$0.4\\
93958 & 11:31:29.21 & +35:09:59.01 & 0.0337 & Comp & 0.30 $\pm$ 0.03 & $<$-1.2 & 9.3 & 147 & 0.97$\pm$0.03 & 3.81$\pm$0.19 & 0.28$\pm$0.00 & 5.8 & 1.38$\pm$0.04 & 22.8$\pm$0.7\\
123656 & 14:20:44.94 & +22:42:36.95 & 0.0307 & AGN & 0.47 $\pm$ 0.07 & $<$-1.1 & 9.3 & 134 & 0.62$\pm$0.03 & 4.01$\pm$0.26 & 0.64$\pm$0.01 & 5.9 & 1.49$\pm$0.02 & 4.0$\pm$0.3\\
\\
\hline
\multicolumn{15}{c}{Gas-Normal}\\
1933 & 13:38:15.39 & -00:23:54.79 & 0.0220 & Comp & 0.27 $\pm$ 0.02 & -0.3 $\pm$ 0.1 & 9.1 & 93 & 0.27$\pm$0.03 & - & 2.02$\pm$0.05 & 1.0 & 1.12$\pm$0.01 & 92.3$\pm$1.2\\
5512 & 23:35:56.57 & +00:42:13.94 & 0.0224 & SF & 0.15 $\pm$ 0.10 & -0.2 $\pm$ 0.1 & 9.3 & 100 & 0.40$\pm$0.03 & 2.17$\pm$0.05 & 5.66$\pm$0.12 & 1.3 & 1.23$\pm$0.10 & 11.9$\pm$1.3\\
47492 & 08:00:24.59 & +25:21:19.87 & 0.0118 & SF & 0.14 $\pm$ 0.02 & -0.2 $\pm$ 0.1 & 8.4 & 54 & 0.14$\pm$0.03 & - & 0.87$\pm$0.03 & 2.0 & 1.06$\pm$0.01 & 89.5$\pm$1.6\\
48538 & 11:33:25.42 & +09:58:59.36 & 0.0210 & SF & 0.19 $\pm$ 0.13 & -0.2 $\pm$ 0.1 & 9.3 & 90 & 0.56$\pm$0.03 & 2.95$\pm$0.23 & 4.25$\pm$0.08 & 0.8 & 1.30$\pm$0.07 & 6.7$\pm$0.8\\
56919 & 15:30:56.86 & +31:36:49.49 & 0.0365 & Comp & 0.33 $\pm$ 0.04 & -0.2 $\pm$ 0.1 & 9.4 & 161 & 0.32$\pm$0.03 & - & 0.25$\pm$0.00 & 6.0 & 1.18$\pm$0.01 & 50.1$\pm$0.8\\
64878 & 09:27:47.64 & +33:12:33.17 & 0.0375 & Comp & 0.26 $\pm$ 0.12 & -0.2 $\pm$ 0.1 & 9.4 & 165 & 0.46$\pm$0.03 & 2.20$\pm$0.09 & 5.18$\pm$0.07 & 1.6 & 1.34$\pm$0.07 & 7.3$\pm$1.0\\
65192 & 09:50:20.09 & +36:04:46.56 & 0.0218 & Comp & 0.22 $\pm$ 0.12 & -0.7 $\pm$ 0.1 & 9.3 & 94 & 0.45$\pm$0.05 & 2.71$\pm$0.14 & 2.96$\pm$0.07 & 1.2 & 1.33$\pm$0.05 & 5.9$\pm$0.9\\
74744 & 10:55:55.13 & +13:56:16.97 & 0.0238 & Comp & 0.29 $\pm$ 0.03 & -0.6 $\pm$ 0.1 & 9.2 & 101 & 0.51$\pm$0.03 & 3.56$\pm$0.17 & 0.47$\pm$0.01 & 2.2 & 1.32$\pm$0.01 & 5.6$\pm$0.3\\
90179 & 10:47:58.08 & +33:55:37.13 & 0.0228 & Comp & 0.24 $\pm$ 0.09 & 0.1 $\pm$ 0.1 & 9.4 & 99 & 0.36$\pm$0.03 & 2.03$\pm$0.05 & 7.82$\pm$0.17 & 1.9 & 1.25$\pm$0.06 & 15.5$\pm$1.4\\
106991 & 10:04:23.34 & +23:13:23.37 & 0.0266 & Comp & 0.30 $\pm$ 0.08 & -0.3 $\pm$ 0.1 & 9.2 & 115 & 0.53$\pm$0.03 & 2.98$\pm$0.20 & 0.36$\pm$0.01 & 6.0 & 1.54$\pm$0.03 & 4.5$\pm$0.3\\
109873 & 08:43:35.13 & +13:03:46.57 & 0.0138 & SF & 0.19 $\pm$ 0.01 & -0.4 $\pm$ 0.1 & 8.8 & 60 & 0.27$\pm$0.03 & 1.88$\pm$0.07 & 1.12$\pm$0.04 & 1.9 & 1.08$\pm$0.01 & 35.8$\pm$0.6\\
114208 & 15:41:24.89 & +11:59:05.48 & 0.0145 & SF & 0.12 $\pm$ 0.02 & -0.6 $\pm$ 0.1 & 9.5 & 60 & 0.79$\pm$0.03 & 3.76$\pm$0.14 & 1.89$\pm$0.07 & 0.7 & 1.34$\pm$0.02 & 17.2$\pm$0.3\\
119809 & 12:24:16.18 & +24:16:01.74 & 0.0289 & SF & 0.18 $\pm$ 0.10 & 0.1 $\pm$ 0.1 & 9.2 & 125 & 0.44$\pm$0.03 & 2.04$\pm$0.13 & 5.27$\pm$0.09 & 3.5 & 0.88$\pm$0.05 & 9.7$\pm$1.2\\
119983 & 12:58:45.57 & +24:14:02.10 & 0.0227 & SF & 0.12 $\pm$ 0.09 & -0.2 $\pm$ 0.1 & 9.4 & 97 & 0.45$\pm$0.03 & 2.28$\pm$0.06 & 5.53$\pm$0.12 & 1.1 & 1.35$\pm$0.06 & 9.8$\pm$0.9\\
124169 & 15:20:14.16 & +17:43:52.95 & 0.0214 & SF & 0.12 $\pm$ 0.01 & -0.1 $\pm$ 0.1 & 9.3 & 93 & 0.23$\pm$0.03 & 1.95$\pm$0.10 & 2.90$\pm$0.07 & 3.4 & 1.12$\pm$0.01 & 105.9$\pm$1.7
\enddata
\tablecomments{Galaxies above the horizontal break are gas-depleted galaxies and galaxies below are gas-normal. Column definitions are: (1) NSAID,  (2) R.A., (3) decl., (4) redshift, (5) BPT categorization from \citet{Reines:2013bp}, (6) BPT distance defined in \S\,2.5, (7) HI residual as defined in \S\,2.3, (8) stellar mass, (9) heliocentric distance, (10) $g-r$ color, (11) NUV-r color, (12) optical effective radius, (13) Sersi\'c index, (14) 4000 \AA break, and (15)  the \Ha\ equivalent width measured from SDSS spectra.}
\label{tbl_int_samples}
\end{deluxetable*}
\end{turnpage}